\documentclass[11pt,a4paper]{article}
\usepackage{ifpdf}               
\usepackage{a4wide}             
\usepackage{amssymb}            
\usepackage[latin1]{inputenc}    
\usepackage[notref,notcite]{showkeys} 
\ifpdf
\usepackage[pdftex]{graphicx}
\usepackage[pdftex,unicode,implicit]{hyperref}
\hypersetup{%
  pdftitle    = {Supersymmetric Coloured/Hairy Black Holes},
  pdfkeywords = {Supergravity, coloured black holes, analytic solution},
  pdfauthor   = {Patrick A.A. Meessen: patrick.meessen@uam.es},
  pdfcreator  = {pdf\LaTeXe\ with package \flqq hyperref\frqq},
  pdfproducer = {pdf\LaTeXe\ with package \flqq hyperref\frqq},
  pdfpagemode = None,  
  pdffitwindow= true,  
  unicode     = true,
  plainpages  = false,
  colorlinks  = true,  
  citecolor   = black,
  urlcolor    = black,
  linkcolor   = black
}
\else
  \usepackage[dvips]{graphicx}
  \usepackage[unicode,implicit]{hyperref}
\fi
\begin{document}
\begin{flushright}
\small
IFT-UAM/CSIC-08-14\\
March 5, 2008
\normalsize
\end{flushright}
\begin{center}
{\large\bf Supersymmetric coloured/hairy black holes}\\[.5cm]
{\bf Patrick~Meessen}\\[.2cm]
{\em Instituto de F\'{\i}sica Te\'orica UAM/CSIC, Facultad de Ciencias C-XVI\\
     C.U. Cantoblanco, E-28049 Madrid, Spain}\\[.5cm]
{\bf abstract}\\
\begin{quote}
{\small 
  We discus all possible spherically symmetric black hole type solutions to an
  $N=2$ supergravity model with $SO(3)$ gauging. The solutions consist of a 
  one parameter family illustrating that the no-hair theorem does not hold and
  an isolated solution that a supersymmetric analogue of a coloured black hole. 
}
\end{quote}
\end{center}
\vspace{.5cm}
Gravity coupled to Yang-Mills fields behaves qualitatively very different from gravity
coupled to Maxwell fields and some of the cherished properties of regular solutions
to Einstein-Maxwell theory fail to hold for EYM theories. Some of these properties
are the Lichnerowicz theorem, which proves the non-existence of non-Minkowski globally
regular solutions in asymptotically flat spacetimes, the no-hair theorem for black holes 
and Israel's theorem that states that a static black hole spacetime is necessarily spherically 
symmetric. Historically, the first of these properties to be seen to fail for EYM was the 
Lichnerowicz theorem: In Ref.~\cite{Bartnik:1988am} Bartnik and McKinnon presented numerical evidence
for the existence of a discrete family of asymptotically flat, spherically symmetric, globally regular 
solutions to $SO(3)$ EYM; furthermore, their solutions evaded the {\em non-Abelian baldness theorem}
\cite{Galtsov:1989ip}, that states that every solution with finite colour charge is an Abelian solution
in disguise, by having no asymptotic colour charge.
Shortly after, these results were generalised in Refs.~\cite{Bizon:1990sr}, 
obtaining a discrete family of asymptotically flat black hole solutions, whence violating the no-hair theorem,
baptising them coloured black holes.
Subsequent research (see {\em e.g.\/} Refs.~\cite{Volkov:1998cc,Winstanley:2008ac}
for more complete information) has not only proven the existence of the aforementioned solutions
but has also shown them to be unstable.\footnote{
  EYM-Higgs models with the Higgs in the adjoint representation are closer to the theories 
  we are going to consider and in those models, stable solutions can be found \cite{Ortiz:1991eu}.
}
\par
In supergravity the above properties also fail to generalise to non-Abelian matter couplings.
For instance,
the stringy generalisation of the Lichnerowicz theorem by Breitenlohner, Maison and 
Gibbons \cite{Breitenlohner:1987dg}, which states that the only asymptotically flat,
globally regular solution of ungauged $d=4$ sugra is Minkowski space, is contradicted
by the Harvey \& Liu monopole in $N=4$ $d=4$ sugra \cite{Harvey:1991jr} or the supersymmetric
embeddings of the 't Hooft-Polyakov monopoles into $N=2$ obtained recently in Ref.~\cite{Huebscher:2007hj}.
The violation of the no-hair conjecture will be illustrated below by a continuous family of black hole
solutions with the same mass, moduli and charges. The non-applicability of Israel's theorem can 
be inferred from Ref.~\cite{Huebscher:2007hj}, as instead of an 't Hooft-Polyakov monopole we could have
embedded an $SO(3)$ multi-monopole, that is not spherically symmetric. The reasoning of why the resulting
spacetime is static is, however, independent of the gauge part, whence a multi-monopole configuration
should exemplify the non-applicability of Israel's theorem.
\par
In order not to clutter this letter with unnecessary technicalities, we shall consider a specific model,
namely the $N=2$ $d=4$ supergravity model based on the special geometry $SU(1,3)/U(3)$. The bosonic
field content of this model consists of the metric, 4 gauge fields $A^{\Lambda}$ ($\Lambda = 0,1,2,3$)
and 3 complex scalar fields $Z^{i}$ ($i=1,2,3$). The general construction of gauged sugras, see {\em e.g.\/}
\cite{Andrianopoli:1996cm}, allows for the gauging of a 4-dimensional subgroup of $SO(1,3)$, but as we are
interested in spherically symmetric solutions, the most interesting gauging is the $SO(3)$ one. 3 of the 
vector fields constitute the gauge fields, leaving $A^{0}$ as an Abelian field; the scalars
transform as a triplet under the $SO(3)$. The bosonic action for the model is
\begin{equation}
  \label{eq:DefAction}
  \int_{4}\sqrt{g}\left[ 
        \textstyle{1\over 2}R
   \ +\ \mathcal{G}_{i\bar{\jmath}}\mathtt{D}_{a}Z^{i}\mathtt{D}\overline{Z}^{\bar{\jmath}}
   \ -\ V(Z,\overline{Z})
   \ +\ \mathrm{Im}(\mathcal{N})_{\Lambda\Sigma} F^{\Lambda}_{ab}F^{\Sigma ab}
   \ -\ \mathrm{Re}(\mathcal{N})_{\Lambda\Sigma} F^{\Lambda}_{ab}\ (\star F)^{\Sigma ab}
  \right]
\end{equation}
In this action the field strengths and the covariant derivatives are given by
\begin{equation}
  \label{eq:DefCovDer}
  F^{0} \ =\ dA^{0} \;\; ,\;\; 
  F^{i} \ =\ dA^{i} +\textstyle{g\over 2}\varepsilon_{jk}{}^{i}A^{j}\wedge A^{k}
  \;\; ,\;\; 
  \mathtt{D}Z^{i} \ =\ dZ^{i} + g\varepsilon_{jk}{}^{i}A^{j}\ Z^{k} \; .
\end{equation}
As the metric $\mathcal{G}$ is K\"ahler it can be derived from a K\"ahler potential, $\mathcal{K}$, which
for the chosen model reads $\mathcal{K} = -\log\left( 1-|Z|^{2}\right)$. 
Please observe that the K\"ahler potential is not only $SO(3)$ invariant, but that it also imposes the 
constraint $0\leq |Z|^{2} \leq 1$; a regular solution must satisfy this bound on its domain of definition,
but as shown in \cite{Bellorin:2006xr}, this is automatically satisfied if the metric is regular. 
The complex
matrix $\mathcal{N}$ can be derived from special geometry, but as it is not explicitly needed, we shall
refrain from writing it down. The potential $V$ is given by
\begin{equation}
   V \; =\; \textstyle{g^{2}\over 4}\ \left( 1 -|Z|^{2}\right)^{-2}\; 
   \left| \left[\ Z\ ,\ \overline{Z}\ \right]\right|^{2} \; ,
\end{equation}
and is positive semi-definite, which means that the black holes will be asymptotically flat.
\par
The static supersymmetric solutions in the timelike class for these type of gauged supergravity models were
recently classified in Ref.~\cite{Huebscher:2007hj}, a complete classification is forthcoming \cite{future},
and we will briefly discuss the structure of these BPS solutions in our model. The static solutions are completely
determined by the real functions $\mathcal{I}^{\Lambda}$ and $\mathcal{J}_{\Lambda}$, defined on 
$\mathbb{R}^{3}$ with coordinates $x^{i}$, that have to satisfy
\begin{equation}\label{eq:DefI}
 \begin{array}{lclclcl}
  \vec{\partial}^{2}\mathcal{I}^{0} & =& 0 &\;\; ,\;\;&
  \mathtt{D}_{i}\mathcal{I} & =& \textstyle{1\over 2}\varepsilon_{ijk}\ \mathcal{F}_{jk} \; ,\\
  & & & & & & \\
  \vec{\partial}^{2}\mathcal{J}_{0} & =& 0 & ,&
  \vec{\mathtt{D}}^{2}\mathcal{J} & =& -g^{2}\ \left[ \mathcal{I},\left[ \mathcal{I},\mathcal{J}\right]\right] \; ,
 \end{array}  
\end{equation}
where we introduced the $\mathfrak{so}(3)$-valued fields $\mathcal{I}$, $\mathcal{J}$ and $\mathcal{F}$.
Furthermore, $\mathcal{F}$ is the fieldstrength for an $SO(3)$ gauge connection defined on $\mathbb{R}^{3}$;
the second equation in Eq.~(\ref{eq:DefI}) is the Bogomol'nyi equation and is the keystone for the construction
of supergravity solutions.
\par
Given a solution to the above equations, the scalars and the metric are given by 
\begin{equation}
  \label{eq:MetDef}
  Z^{i} \; =\; \frac{ i\mathcal{I}^{i}\ -\ \mathcal{J}_{i} }{ \mathcal{J}_{0} \ +\ i\mathcal{I}^{0} } 
  \;\; ,\;\;
  ds^{2} \; =\; 2|X|^{2}\ dt^{2} \; -\; \textstyle{1\over 2|X|^{2}}d\vec{x}^{2}\; ,
\end{equation}
and the factor determining the metric is
\begin{equation}
  \label{eq:MetFact}
  \textstyle{1\over 2|X|^{2}} \; =\; \left( \mathcal{I}^{0}\right)^{2}
        \ +\ \left( \mathcal{J}_{0}\right)^{2}
        \ -\ \left(\vec{\mathcal{I}}\right)^{2} 
        \ -\ \left(\vec{\mathcal{J}}\right)^{2}   \; .
\end{equation}
As was pointed out in Ref.~\cite{Huebscher:2007hj}, staticity imposes the constraint
$\mathcal{J}_{\Lambda}\vec{\mathtt{D}}\mathcal{I}^{\Lambda}=\mathcal{I}_{\Lambda}\vec{\mathtt{D}}\mathcal{J}^{\Lambda}$
and we will obviate this constraint by taking $\mathcal{J}_{\Lambda}=0$, even though the general case is a trivial extention of what follows. This means that we will be dealing with purely magnetic solutions.
The remainder of this letter, then, is to find all the acceptable spherically symmetric solutions to the 
Bogomol'nyi equation and to use them to construct black holes. 
\par
Since we are dealing with a non-Abelian theory, we have to deal with the problem of giving a sensible 
definition of the charges:
a way to get over this problem is to impose a gauge-fixing, see {\em e.g.\/} \cite{Goddard:1977da,Volkov:1998cc}, 
which for spherical symmetry means making the standard hedgehog Ansatz for $(A,\mathcal{I})$, namely
\begin{equation}
  \label{eq:ColAnsatz}
  \mathcal{I}^{i} \; =\; H(r)\ x^{i} \;\; ,\;\; 
  A_{m}^{i} \; =\; -\varepsilon_{mn}{}^{i}\ x^{n}\ A(r) \; .
\end{equation}
In terms of the functions $A$ and $H$ the asymptotic behaviour for finite energy solutions is
\begin{equation}
  \label{eq:8}
  \lim_{r\rightarrow\infty} A \ =\ -\mathsf{P}\ r^{-2} \ +\ldots \;\;\; ,\;\;\; 
  \lim_{r\rightarrow\infty} H \ =\ \phi\ r^{-1} \; +\; \mathsf{P}\ r^{-2} \ +\ldots \;\; ,
\end{equation}
where $\mathsf{P}$ is called the magnetic colour charge and $\phi$ sets the scale
of the asymptotic gauge symmetry breaking as $\lim_{r\rightarrow\infty}\vec{\mathcal{I}}^{2} = \phi^{2}$.
Plugging the above Ansatz into the Bogomol'nyi equation one finds
\begin{eqnarray}
  \label{eq:ColHof1}
  r\partial_{r}\left[ A +H\right] \ =\ gr^{2}\ A\ \left[ A + H\right] \;\;\; ,\;\;\;
  r\partial_{r}A \ +\ 2A \ =\ H\ \left( 1 + gr^{2}A\right) \; .
\end{eqnarray}
By changing coordinates to $e^{2\xi}=gr^{2}$ and redefining the functions by $I= gr^{2}\ H$
and $N= gr^{2}A + 1$, we find a set of autonomous first order differential equations, namely
\begin{equation}
  \label{eq:1}
  \partial_{\xi}I \ =\ N^{2} \ +\ I -1 \;\; ,\;\;  \partial_{\xi}N \ =\ I\ N \; , 
\end{equation}
\begin{figure}
\centering
\includegraphics[height=5cm]{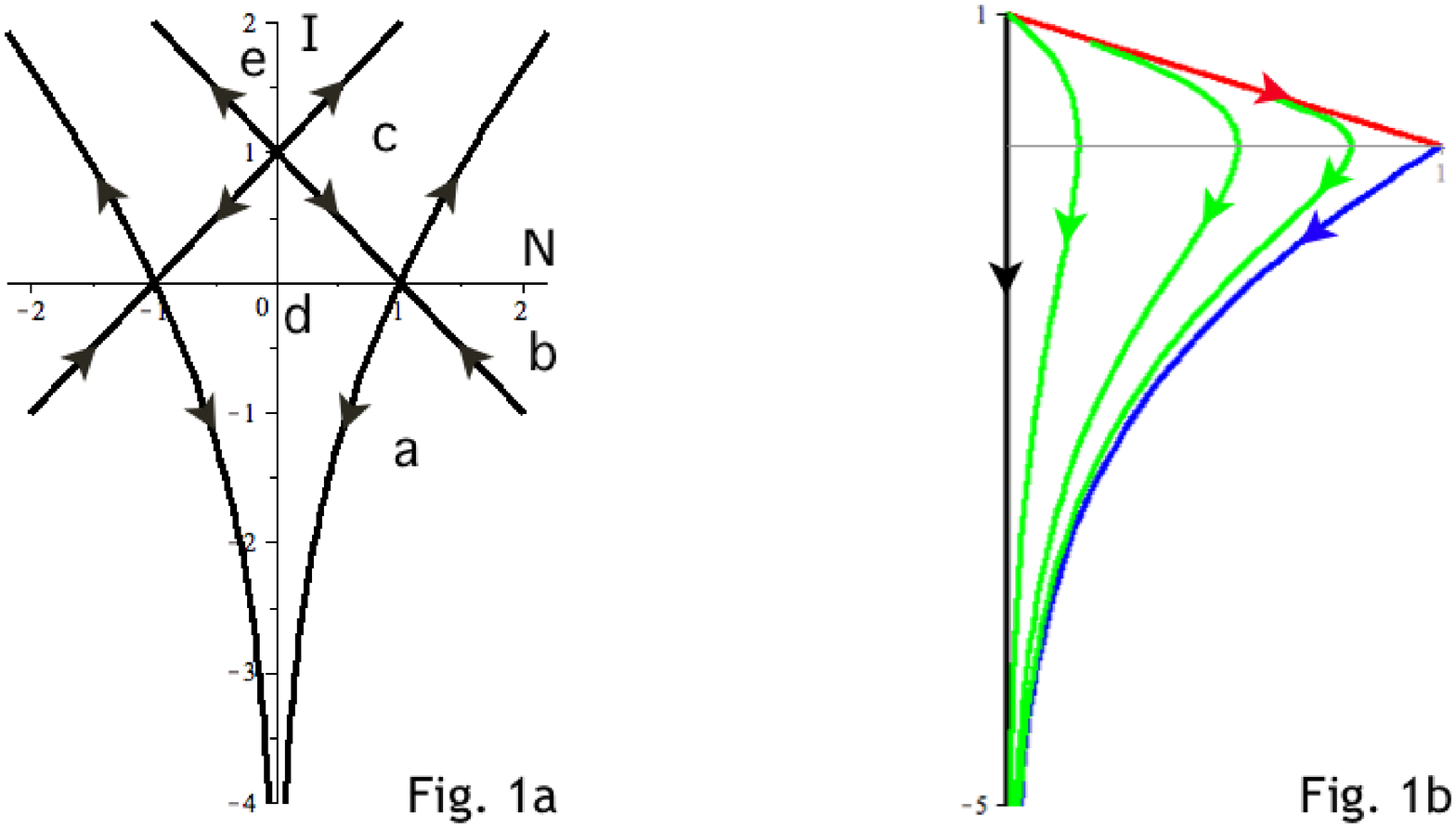}
\caption{\label{fig:Flow} Figure 1a shows the separatrices and fixed points of the system (\ref{eq:1}).
 Figure 1b shows the flows in the region $d$; the closer a flowline is to the line $N=0$ the greater is
 $s$, the Protogenov hair.
}
\end{figure}
The flow of this system was analysed by Protogenov \cite{Protogenov:1977tq} and is depicted in
Fig.~(\ref{fig:Flow}): the lines correspond to separatrices, {\em i.e.\/}~the frontiers of regions 
with different dynamics, the arrows indicate the direction of the flow and the fixed points of
the system are located at the intersections of the separatrices. The desired asymptotic behaviour in
Eq.~(\ref{eq:1}) translates to the fact that the admissible flows should asymptote to $N=cnst.$ when
$\xi\rightarrow\infty$. This excludes regions $b$ and $c$ and almost
all of region $e$: as is easy to be seen, a generic flow in the region $e$ is pushed off towards infinity,
the only exception being the flow with $N(\xi )=0$. The flows in region $a$ are well behaved at late times
in the flow, but start out by blowing up which means that they cannot be used to build a regular solution.
The flows in the central diamond, region $d$, are perfectly well behaved and give rise to hairy black holes.
\par
The system (\ref{eq:1}) is invariant under the interchange $N\leftrightarrow -N$, so that we will concentrate
on the right side of the diamond ({\em i.e.\/} $N\geq 0$): figure (1b) depicts the possible flows in the diamond.
The first remark about the flows in the diamond is that they correspond
to complete trajectories, by which we mean that the starting point is $\xi=-\infty$ ($r=0$) and the endpoint $\xi=\infty$
($r=\infty$). The fact that the trajectories
flow towards $N=0$ then means that the solutions have the correct asymptotic behaviour with magnetic colour charge $\mathsf{P}=1$.
Of course, the beginning of a given flow is also important and there are 2 such points, namely
$(I,N)=(1,0)$ and $(I,N)=(0,1)$. As already observed by Protogenov, the separatrix between the regions $a$ and $d$ corresponds to the
't Hooft-Polyakov monopole in the Bogomol'nyi-Prasad-Sommerfield limit, {\em i.e.\/}
\begin{equation}
  \label{eq:2}
  H \; =\; \frac{1}{g r^{2}}\left[ 1\ -\ \mu r\coth\left( \mu r\right)\right] \;\; ,\;\;
  A \; =\; \frac{1}{g r^{2}}\left( \mu r\sinh^{-1}(\mu r) \ -\ 1\right)\; ,
\end{equation}
where $\mu$ is a positive constant that is proportional to the mass of the W-bosons arising outside the
core of the monopole. 
As was pointed out in Ref.~\cite{Huebscher:2007hj}, this solution together with a constant $\mathcal{I}^{0}$
leads to a globally regular spacetime. The fact that we can take $\mathcal{I}^{0}$ to be constant 
is not only due to the fact that the solution
is regular for all $r$, but most of all due to the fact that the beginning of flowline, $(0,1)$, corresponds
to the region in spacetime where the Higgs field and the connection vanish, or to put it differently: it is the point of gauge symmetry
restoration. One can of course defrost the harmonic function corresponding to the graviphoton and find the intersection of an 
extreme Reissner-Nordstrom black hole with an 't Hooft-Polyakov monopole.
The $N=0$ trajectory corresponds to what the authors of Ref.~\cite{Huebscher:2007hj} called a {\em black hedgehog},
\begin{equation}
  \label{eq:3}
  H \; =\; \frac{1}{r}\left[ -\mu g^{-1} \ +\ \frac{1}{gr}\right] \; ,\;
  A \; =\; -\frac{1}{gr^{2}} \; .
\end{equation}
Motion in the diamond has $\mu>0$ but as discussed above, the motion along the positive $I$-axis is well-defined, and this is given 
by $\mu <0$.
\par
The solutions for the flowlines inside the half-diamond read \cite{Protogenov:1977tq}
\begin{equation}
  \label{eq:4}
  H \; =\; \frac{1}{g r^{2}}\left[ 1\ -\ \mu r\coth\left( \mu r+s\right)\right] \;\; ,\;\;
  A \; =\; \frac{1}{g r^{2}}\left( \mu r\sinh^{-1}(\mu r +s) \ -\ 1\right)\; ,
\end{equation}
where $s\in\mathbb{R}^{+}$ is a parameter we call the {\em Protogenov hair}.\footnote{
  One can also consider the case $s<0$, which leads to motions in the regions $a$ and $e$. More to the 
  point, solutions with $s<0$ and $r>-s$ cover region $a$, and solutions with $s<0$ and $r\in (0,-s)$ cover all
  of region $e$.
} 
It is clear that the 't Hooft-Polyakov monopole
corresponds to $s=0$, which remarkably is a regular limit in the solution but is singular in the flow. On the other extreme
we have the limit $s\rightarrow\infty$: on the solutions this limit is taken with $r$ fixed and the result is the black hedgehog
(\ref{eq:3}) and in the flow diagram nothing special happens. At this point it is worth to observe that the behaviour of the $s\neq 0$
solutions in their asymptotic regimes, {\em i.e.\/} $\xi\rightarrow \mp\infty$ is universal. The $\xi\rightarrow\infty$ universality
is due to the above mentioned attractive character of the $N=0$ line for the flows in the lower half plane.
The universality in the $\xi\rightarrow -\infty$ is due to the fact that the point $(I,N)=(1,0)$ is an unstable fixed point.
As far as the construction of black holes is
concerned, it is this universality that guarantees that we can use the solution (\ref{eq:4}) to construct supersymmetric hairy black holes.
\par
Substituting Eq.~(\ref{eq:4}) into Eq.~(\ref{eq:MetFact}) and taking $\mathcal{I}^{0}$ to be an arbitrary
spherically symmetric harmonic function, we see that the metrical factor is
\begin{equation}
  \textstyle{1\over 2|X|^{2}} \; =\; \left( h \ +\ \frac{p}{r}\right)^{2}
       \ -\ \frac{\mu^{2}}{g^{2}}\left(
          \coth (\mu r+s) \ -\ \frac{1}{gr}
       \right)^{2} \; .
\end{equation}
It is clear that this metrical factor is well-behaved when $r>0$ and that it asymptotes to a constant: normalising
the asymptotic behaviour to that of Minkowski space then means that $h^{2}=1+\mu^{2}g^{-2}$.
Moreover, as $\coth$ reaches its asymptotic value exponentially it does not contribute to the asymptotic mass, which
is easily calculated to be $M= hp + \mu g^{-1}$ and must be positive for a regular solution. The last ingredient
needed to show that the constructed spacetime is a black hole is the presence of a horizon located at $r=0$.
This imposes the constraint that the 'entropy´ be positive, 
\begin{equation}
  \label{eq:7}
  S_{bh} \; =\; p^{2} \ -\ \frac{1}{g^{2}} \; > 0\; .
\end{equation}
For a given $\mu$ and $g$ we can always choose $p$ and the
sign of the modulus $h$ such that the mass and the entropy are positive. Furthermore, as the resulting black hole
spacetimes are not uniquely specified by the asymptotic charges, moduli and the mass, they illustrate the 
break down of the no-hair conjecture.
\par
At this point the qualitative properties of the separatrix between the regions $c$ and $d$ (see Fig. (1b)) should be more than obvious: 
firstly, the flow starts
at the same point as the black hedgehog, whence there is no obstruction to build a black hole spacetime. Secondly, the flow ends
at the point $(I,N)=(0,1)$, the point of gauge-symmetry restoration, so that the solution has asymptotically vanishing Higgs field
and magnetic charge! Thirdly, the separatrix corresponds to a regular flow as is obvious from the solution \cite{Protogenov:1977tq}\footnote{
  The separatrix between regions $a$ and $b$ also `touches' the point $(0,1)$, but cannot be used as the flow
  is singular. At the level of the solution this means that $H$ blows up at some finite $r>0$, which cannot be
  compensated for.
}
\begin{equation}
  \label{eq:5}
  H \; =\; -A \; =\; \frac{1}{gr^{2}}\left(\frac{1}{1+\lambda^{2}\ r}\right) \; .
\end{equation}
The metrical factor for this case can be written as
\begin{equation}
  \label{eq:6}
  \frac{1}{2|X|^{2}} \; =\; (h+\textstyle{p\over r})^{2} \ -\ \frac{1}{g^{2}r^{2}}\left( \frac{1}{1+\lambda^{2}r}\right)^{2} \; .
\end{equation}
Imposing asymptotic flatness fixes $h=\pm 1$, which means that the mass of the object is $M= hp$. The positivity of the mass
can always be guaranteed by choice of the sign of $h$. The last ingredient we need is the presence
of a horizon located at $r=0$. Universality implies that the entropy is given 
by Eq. (\ref{eq:7}), so that as long as $M > g^{-1}$ we have a well defined black hole spacetime.
\par
The coloured black holes in EYM theory have the interpretation of being made up of a Schwarzschild black hole and a 
Bartnik-McKinnon particle \cite{Volkov:1998cc}, and it is tempting to look for an analogue of the BM-particle,
{\em i.e.\/} a globally regular solution whose colour charge vanishes asymptotically.
Such a solution, however, does not exist as it would require a regular flow beginning at $(I,N)=(0,\pm 1)$ and flowing towards
$(0,1)$ or $(0,-1)$; it is clear from the flow diagram (1a) that such flows do not exists.
\par
The black hole we found can best be described as an extreme Reissner-Nordstrom black hole covered with 
a coloured cloud and the natural question is whether there are $N=2$ sugra models in which one can embed
a coloured cloud without the need for an extreme RN black hole. 
The solutions presented in this letter can also be embedded into sugras with a more direct link to string
theory. These embeddings should shed some light on the strange characteristics of the solutions or why string theory
chooses not to allow them.
\section*{Acknowledgements}
The author wishes to thank R. Emparan, M. H\"ubscher, D. Klemm, J. de Medinaceli, D. Mansi, S. Vaul\`a and 
especially T. Ort\'{\i}n for fruitful discussions.
This work was partially supported by the Spanish MEC grants 
FPA2006-00783, the CAM grant HEPHACOS P-ESP-00346, by the EU RTN grant
MRTN-CT-2004-005104, the Spanish Consolider-Ingenio 2010 program CPAN
CSD2007-00042 and mostly by the {\em Fondo Social Europeo} through an
I3P-doctores scholarship.

\end{document}